\documentclass[]{article}

\usepackage{graphicx}

\usepackage{texdraw}
\usepackage{tikz}
\usepackage{caption}
\usepackage{epic,eepic}
\usepackage{amsfonts}
\usepackage{amssymb}
\usepackage{color}

\newtheorem{rem}{Remark}[section]
\newcommand{\br}{\begin{rem}}
\newcommand{\er}{\end{rem}}
\newtheorem{ex}{Example}[section]
\newcommand{\bex}{\begin{ex}}
\newcommand{\eex}{\end{ex}}
\newtheorem{Def}{Definition}[section]
\newcommand{\bd}{\begin{Def}}
\newcommand{\ed}{\end{Def}}
\newtheorem{theorem}{Theorem}[section]
\newcommand{\bt}{\begin{theorem}}
\newcommand{\et}{\end{theorem}}
\newtheorem{corollary}{Corollary}[section]
\newcommand{\bc}{\begin{corollary}}
\newcommand{\ec}{\end{corollary}}
\newtheorem{lemma}{Lemma}[section]
\newcommand{\bl}{\begin{lemma}}
\newcommand{\el}{\end{lemma}}
\newcommand{\be}{\begin{equation}}
\newcommand{\ee}{\end{equation}}
\newcommand{\bea}{\begin{eqnarray}}
\newcommand{\eea}{\end{eqnarray}}

\newtheorem{prop}{Proposition}[section]
\newcommand{\bpr}{\begin{prop}}
\newcommand{\epr}{\end{prop}}
\newcommand{\ds}{\displaystyle}
\newtheorem{proof}{Proof}[section]
\newcommand{\bpf}{\begin{proof}}
\newcommand{\epf}{\end{proof}}

\begin{document}
\title{$2^n-$rational maps}
\author{
Pavlos Kassotakis
\thanks{\emph{Present address:} Department of Mathematics and Statistics University of Cyprus,
P.O Box: 20537, 1678 Nicosia, Cyprus;
\newline \emph{e-mails:} {\tt  pavlos1978@gmail.com}}
\and
Maciej Nieszporski
\thanks{\emph{Present address:} Katedra Metod Matematycznych Fizyki, Wydzia\l{} Fizyki, Uniwersytet Warszawski,
ul. Pasteura 5, 02-093 Warszawa, Poland;
\emph{e-mail:} {\tt maciejun@fuw.edu.pl}}
\and
Pantelis Damianou
\thanks{\emph{Present address:} Department of Mathematics and Statistics University of Cyprus,
P.O Box: 20537, 1678 Nicosia, Cyprus;
\newline \emph{e-mail:} {\tt  damianou@ucy.ac.cy}}
}
\maketitle

\begin{abstract}
\noindent We present a natural extension of the notion of {\it nondegenerate rational maps} ({\it quadrirational maps}) to arbitrary dimensions. We refer to these maps as {\it $2^n-$rational} maps. In this note we construct a rich family of $2^n-$rational maps. These maps by construction are involutions and highly symmetric in the sense that the maps and their companion maps have the same functional form.
\end{abstract}
\section{Introduction}

In \cite{et-2003} Etingof introduced the notion of {\it nondegenerate rational maps}. These maps arose in the interplay between studies on {\it set-theoretical solutions of the quantum Yang-Baxter equation} \cite{yang-1967,baxter-1970} and the theory of geometric crystals \cite{berenstein-1992}.  Set-theoretical solutions of the quantum Yang-Baxter equation were introduced in \cite{sklyanin-1988,drin}. In \cite{VeselovYB}, among various connections with integrability, the name {\it Yang-Baxter maps} was proposed instead of  set-theoretical solutions. Also, instead of the term  nondegenerate rational maps,  the name {\it quadrirational maps} was coined in \cite{ABSf}. We find the terminology of  quadrirational maps more adequate for this note, hence we will use it from now on. In recent years many results on the connection between  quadrirational Yang-Baxter maps and the theory of discrete integrable systems  were obtained \cite{NY-1998,KNY-2002A,pap2-2006,PSTV,KaNie,PKMN2,KaNie5,atk-2012,atk-2013,atkinson-2014,sokor-2012,koul-2011,sergeev-2015}.

  A rational map  $R: \mathbb{CP}^1\times \mathbb{CP}^1 \ni (x,y) \mapsto    (X,Y) \in \mathbb{CP}^1\times \mathbb{CP}^1$ is called  quadrirational if the maps $R^{-1}: (X,Y)\mapsto (x,y),$ $R^c: (X,y)\mapsto (x,Y),$ and $(R^c)^{-1}: (x,Y)\mapsto (X,y)$ are rational maps as well.  Here, $R^{-1}$ is  the inverse of the map $R$ and  $R^c$    the so-called {\it companion map} of $R$. If all maps and the companion maps have the same functional form, then we refer to it as  quadrirationality in the  narrow sense.  This note is about  quadrirationality and moreover on its generalisation to $n$ dimensions, {\it $2^n-$rationality} (see Definition 3.1), mainly in the  narrow sense.

The QRT map was introduced by Quispel, Roberts and Thompson \cite{qrt1} as a family of integrable maps on the plane. In \cite{tsuda-2004} Tsuda derived the conditions that led to periodic QRT maps. Although in \cite{tsuda-2004} there is no mention of  quadrirationality, it turns out that the period $2$ QRT map is a quadrirational one in the  narrow sense. An alternative way to introduce  quadrirationality via factorisation of involutions was presented in \cite{me-nalini} (see also \cite{kas-2010}).

In Section \ref{sec2}, we are recalling  the QRT construction and we present some  quadrirational subclasses of it. In Section \ref{sec3} we present a QRT generalisation  which was introduced in \cite{tsuda-2004}. Next we consider a natural extension of  quadrirationality to arbitrary dimensions which we refer to as  $2^n-$rationality. We arrive to an explicit form of   $2^n-$rational maps in the  narrow sense.  Afterwards we present in detail the $n=2$ and $n=3$ cases i.e.  quadrirational and  octarational maps respectively. Finally, in Section \ref{sec4} we end this article with some conclusions and perspectives.

\section{Quadrirational QRT Maps} \label{sec2}
The QRT map  \cite{qrt1} $\phi: \mathbb{CP}^1\times \mathbb{CP}^1 \ni (x,y)\mapsto (X,Y)\in \mathbb{CP}^1\times \mathbb{CP}^1$ is defined by the composition of two non-commuting involutions $\sigma^1,$ $\sigma^2,$  which both preserve the same  invariant
\[
{\ds H(x,y)=\frac{{\bf X}^TA_0{\bf
Y}}{{\bf X}^TA_1{\bf Y}}},
\]
where ${\bf X}$, ${\bf Y}$ are vectors ${\bf X}=(x^2,x,1)^T,$ ${\bf Y}=(y^2,y,1)^T$ and $A_0,$ $A_1$ are two  $3\times 3$ matrices,
\[
{\ds A_i=\left(\begin{array}{ccc}
\alpha_i& \beta_i &\gamma_i\\
\delta_i& \epsilon_i& \zeta_i\\
\kappa_i& \lambda_i & \mu_i
\end{array}\right)}.
\]
The QRT is the composition $\sigma^2\circ \sigma^1$ of the non-commuting involutions $\sigma^1,$ $\sigma^2.$ The involution $\sigma^1$ is obtained from the solution of the equation $H(X,y)-H(x,y)=0$ for $X$ and the involution $\sigma_2$ from the solution of  $H(x,Y)-H(x,y)=0$  for $Y.$  Each of these  equations  has 2 solutions. One solution is the identity map and  by choosing the other one we arrive at
\be \label{QRT}
\sigma^1: (x,y)\mapsto (X,Y)= \left({\ds \frac{f_1(y)-f_2(y)x}{f_2(y)-f_3(y)x}},y\right), \quad
\sigma^2: (x,y)\mapsto (X,Y)=\left(x, {\ds\frac{g_1(x)-g_2(x)y}{g_2(x)-g_3(x)y}}\right)
    \ee where  $$
\begin{array}{c}
{\ds (f_1(y),f_2(y),f_3(y))^T=(A_0{\bf Y})\times(A_1{\bf Y})},\\[3mm]
{\ds  (g_1(x),g_2(x),g_3(x))^T=(A_0^T{\bf X})\times(A_1^T{\bf X})}.
    \end{array}
$$
To recapitulate, the QRT map is defined by the equations
\be \label{dqrt}
H(x,y)=H(X,y)=H(X,Y).
\ee
The QRT map preserves a linear pencil of biquadratics curves $h(x,y;t)={\bf X}^TA_0{\bf
Y}-t{\bf X}^TA_1{\bf Y}=0.$
Base points of a linear pencil of biquadratic curves, are the points which are contained in all curves of the linear pencil.  They are determined by
\be \label{base-points}
x=\frac{f_1(y)}{f_2(y)}=\frac{f_2(y)}{f_3(y)},\quad \mbox{or} \quad y=\frac{g_1(x)}{g_2(x)}=\frac{g_2(x)}{g_3(x)}.
\ee
For a given  bi-quadratic polynomial, quantities that are covariant under M\"{o}bius transformations action on the bi-quadratic under consideration are called relative invariants of the latter. They are given by the formulas
$$
\begin{array}{l}
i_2(h,x,y)=2h h_{xxyy}-2h_x h_{xyy}-2h_y h_{yxx}+2h_{xx}h_{yy}+h_{xy}^2,\\ [3mm]
i_3(h,x,y)=\frac{1}{4}\det\left(\begin{array}{ccc}
                      h&h_{x}&h_{xx}\\
                      h_y&h_{xy}&h_{xxy}\\
                      h_{yy}&h_{xyy}&h_{xxyy}
                      \end{array}\right)=\det(A_0-t A_1).
\end{array}
$$

\bpr \label{prop1}
For $i_3=0$ the associated QRT involutions $\sigma^1, \sigma^2$ commute  and the QRT map is an involution.
\epr

\bpf
There is
$$\begin{array}{ll}
i_3=\det(A_0-t A_1)=&{\ds  -t^3 \det A_1+t^2 \left(\frac{\det(A_0+A_1)+\det(A_0-A_1)}{2}-\det A_0\right)}\\ [3mm]
&{\ds +t \left(\frac{\det(A_0-A_1)-\det(A_0+A_1)}{2}-\det A_1\right)+\det A_0=0}.
\end{array}
$$
 Demanding the coefficients of this polynomial on $t$ to be zero we obtain:
\be \label{cond1}
\det A_0=\det A_1=0=\det(A_0-A_1)=\det(A_0+A_1).
\ee
Note that $i_3$ is by definition covariant under M\"{o}bius transformations. By a certain M\"{o}bius transformation we can send 2 of the base points of the linear pencil to $(0,0)$ and $(\infty, \infty).$
Then the  parameters matrices have the form: ${\ds A_i=\left(\begin{array}{ccc}
0& \alpha_i &\zeta_i\\
\beta_i& \epsilon_i& \gamma_i\\
\eta_i& \delta_i &0
\end{array}\right)} \; \mbox{i}=0,1$ and the conditions (\ref{cond1}) are exactly the conditions obtained by Tsuda (see Theorems 2.4 and 3.4 in \cite{tsuda-2004}) for the QRT to be periodic of period 2, from which it follows that the QRT involutions commute. Hence for  $i_3=0$ the associated QRT involutions $\sigma^1, \sigma^2$ commute  and the QRT map is an involution.

\epf
From now on we focus on  the following  solution of the conditions (\ref{cond1}):
\be \label{tma}
{\ds A_i=\left(\begin{array}{ccc}
0& \alpha_i &0\\
\beta_i& \epsilon_i& \gamma_i\\
0& \delta_i &0
\end{array}\right)},\;\; i=0,1.
 \ee
   The QRT map for this choice of parameter matrices, was given explicitly  in \cite{tsuda-2004}. In the following Proposition  we repeat a result of \cite{tsuda-2004} and we prove our observation i.e. in this case the period 2  QRT map is a  quadrirational map.

\bpr \label{prop2}
The QRT map associated to the integral
\be \label{tqrt}
H(x,y)=\frac{\alpha_0 +\alpha_1 x+\alpha_2y+\frac{\beta_1}{x}+\frac{\beta_2}{y} }
{\gamma_0 +\gamma_1 x+\gamma_2y+\frac{\delta_1}{x}+\frac{\delta_2}{y}}
\ee
reads
\be \label{q-qrt}
\phi: (x,y)\mapsto (X,Y)=\left(\frac{1}{x}\frac{\beta_1-\delta_1 H(x,y)}{\alpha_1-\gamma_1 H(x,y)},\frac{1}{y}\frac{\beta_2-\delta_2 H(x,y)}{\alpha_2-\gamma_2 H(x,y)} \right).
\ee
It is an involution and moreover a  quadrirational map in the  narrow sense i.e. the map the companion map and their inverses have the same functional form.
\epr

\bpf
It is easy to show that the parameter matrices $A_i$ associated to the  integral (\ref{tqrt}) are of the form (\ref{tma}), so $i_3=0$  and the QRT map is an involution.
For the integral (\ref{tqrt}), the QRT map is $\phi: (x,y)\mapsto(X,Y)$ where:
\be \label{qyb}
 \begin{array}{l}
X={\ds \frac{(\beta_1\delta_0-\beta_0\delta_1)y^2+(\epsilon_1\delta_0-\epsilon_0\delta_1)y+\gamma_1\delta_0-\gamma_0\delta_1-(\alpha_0\delta_1-\alpha_1\delta_0)yx}
{(\alpha_0\delta_1-\alpha_1\delta_0)y-\left((\alpha_1\beta_0-\alpha_0\beta_1)y^2+(\alpha_1\epsilon_0-\alpha_0\epsilon_1)y+\alpha_1\gamma_0-\alpha_0\gamma_1\right)x}}\\[3mm]
  Y={\ds \frac{(\alpha_1\gamma_0-\alpha_0\gamma_1)x^2+(\epsilon_1\gamma_0-\epsilon_0\gamma_1)x+\gamma_0\delta_1-\gamma_1\delta_0-(\beta_0\gamma_1-\beta_1\gamma_0)xy}
  {(\beta_0\gamma_1-\beta_1\gamma_0)x-\left((\alpha_0\beta_1-\alpha_1\beta_0)x^2+(\beta_1\epsilon_0-\beta_0\epsilon_1)x+\beta_1\delta_0-\beta_0\delta_1\right)y}}.
    \end{array}
\ee
 In terms of the integral $H,$ mapping  (\ref{qyb}) simplifies to
 \be \label{qybq}
 \phi: (x,y)\mapsto (X,Y)=\left(\frac{1}{x}\frac{\beta_1-\delta_1 H(x,y)}{\alpha_1-\gamma_1 H(x,y)},\frac{1}{y}\frac{\beta_2-\delta_2 H(x,y)}{\alpha_2-\gamma_2 H(x,y)} \right).
 \ee

The companion map $\phi^c$ of the mapping $\phi$, is the map $\phi^c: \; (X,y)\mapsto (x,Y)$ i.e.  from  (\ref{qybq}) we need to solve for $x, Y$ in terms of $X, y$. From the first equality of (\ref{qybq}) we obtain
$
{\ds x X= \frac{\delta_0-\delta_1 H(x,y)}{\alpha_0-\alpha_1 H(x,y)}},
$
from (\ref{dqrt}) we have   $H(x,y)=H(X,y),$  so
 ${\ds x= \frac{1}{X}\frac{\delta_0-\delta_1 H(X,y)}{\alpha_0-\alpha_1 H(X,y)}},$ and  from the second equality of (\ref{qybq})  ${\ds Y=\frac{1}{y}\frac{\gamma_0-\gamma_1 H(X,y)}{\beta_0-\beta_1 H(X,y)}}.$ So the companion map  is
$$
\phi^c: \; (X,y)\mapsto  (x,Y)=\left(\frac{1}{X}\frac{\delta_0-\delta_1 H(X,y)}{\alpha-\alpha_1 H(X,y)},\frac{1}{y}\frac{\gamma_0-\gamma_1 H(X,y)}{\beta_0-\beta_1 H(X,y)} \right).
$$
The map $\phi^c$   has the same functional form as $\phi$ so it is also an involution.
 Therefore all the 4 maps $\phi, \phi^{-1}, \phi^c, (\phi^c)^{-1}$ have the same functional form and the map $\phi$ is  quadrirational in the  narrow sense.

\epf

\section{$2^n-$rational maps} \label{sec3}
Generalisations of the QRT map have been proposed by various authors \cite{capel-2001,iatrou2003,tsuda-2004,qrt-2006,kassotakis-2006,Adler-2006}. In order to derive $2^n-$rational maps we focus on the  generalisation given in \cite{tsuda-2004} which includes the generalisation given in \cite{capel-2001}.
Namely, a map $\phi: (\mathbb{CP}^1)^n \ni (x^1,x^2, \ldots, x^n)\mapsto (X^1,X^2,\ldots, X^n)\in (\mathbb{CP}^1)^n$  that  is defined by the composition of $n$ non-commuting involutions $\sigma^1,$ $\sigma^2,\ldots \sigma^n,$ i.e. which  preserve the same  integral
\be \label{tsuda-integral}
 H(x^1,x^2,\ldots x^n)=\frac{{\ds\sum_{j_1,j_2,\ldots, j_n =0}^2\alpha_{j_1,j_2,\ldots, j_n}(x^1)^{2-j_1}(x^2)^{2-j_2}\ldots (x^n)^{2-j_n}}}
{{\ds\sum_{j_1,j_2,\ldots, j_n =0}^2\beta_{j_1,j_2,\ldots, j_n}(x^1)^{2-j_1}(x^2)^{2-j_2}\ldots (x^n)^{2-j_n}}}.
\ee
 Note that with $(\mathbb{CP}^1)^n$ we denote $(\mathbb{CP}^1)^n:=\underbrace{\mathbb{CP}^1\times
\mathbb{CP}^1\times \ldots \times \mathbb{CP}^1}_{n \; {\tiny \mbox{times}}}.$

The involution $\sigma^k$  is defined by the non-trivial solution of $H(x^1,x^2,\ldots X^k, \ldots, x^n)= H(x^1,x^2,\ldots x^n),$ for $X^k.$ Similarly for the other $n-1$ involutions. From now on we will call mapping $\phi$ as {\it Tsuda} map.
Note that for $n=2$ we have exactly the QRT map. To recapitulate, the Tsuda map is defined by the equations:
\be \label{ctsuda}
\forall m\in \{1\ldots n\}, \;\; H(x^1,\ldots x^{m-1},x^m, x^{m+1},\ldots x^n)=H(X^1,\ldots X^{m-1},X^m, x^{m+1},\ldots x^n)
\ee
and preserves the linear pencil of manifolds

\be \label{n-quadratic}
\begin{array}{l}
h(x^1,x^2,\ldots x^n;t)={\ds\sum_{j_1,j_2,\ldots, j_n =0}^2\alpha_{j_1,j_2,\ldots, j_n}(x^1)^{2-j_1}(x^2)^{2-j_2}\ldots (x^n)^{2-j_n}}-\\ [3mm]
t\;{\ds\sum_{j_1,j_2,\ldots, j_n =0}^2\beta_{j_1,j_2,\ldots, j_n}(x^1)^{2-j_1}(x^2)^{2-j_2}\ldots (x^n)^{2-j_n}}=0.
\end{array}
\ee
 For $n=2$ the pencil of manifolds is a linear pencil of elliptic curves. For $n=3$ we have a linear pencil of K-3 surfaces and for $n \geq 4$ we have  Calabi-Yau manifolds.

Before we present the definition of a $2^n-$rational map, we introduce some notation.  Let $M$ the set of $n$ indices, $M:=\{1,\ldots n\}$ and  $m$ any subset of the set $M$ i.e. $m:=\{m_1,m_2,\ldots m_k\}, \; k\leq n.$ Also $m^c$ the complement of the set $m$ so  $m^c:=M \setminus m$.  Finally, with ${\mathcal X}^{m}$ and ${\mathcal X}^{m^c}$ we denote the coordinates ${\mathcal X}^{m}:=(x^1,x^2,\ldots, X^{m_1}, x^{m_1+1},\ldots, X^{m_2},\ldots X^{m_k},\ldots x^n )$ and  ${\mathcal X}^{m^c}:=(X^1,X^2,\ldots, x^{m_1}, X^{m_1+1},\ldots, x^{m_2},\ldots x^{m_k},\ldots X^n ).$ Now we are ready to give the following definition.

\bd
 A rational map  $R:\; (\mathbb{CP}^1)^n \ni (x^1,x^2,\ldots, x^n) \mapsto    (X^1,X^2,\ldots, X^n) \in (\mathbb{CP}^1)^n$ is a $2^n-$rational map if  all $2^n$ maps
 $R^{c_m}: (\mathbb{CP}^1)^n \ni {\mathcal X}^{m}\mapsto {\mathcal X}^{m^c}\in (\mathbb{CP}^1)^n,$ are rational maps. The maps $R^{c_m}$ are called companion maps of $R$.
 \ed

\bpr
The Tsuda map $\phi$ associated to the integral
\be \label{tsuda-int}
H(x^1, \ldots , x^n)=\frac{\alpha_0+\sum_{i=1}^n \alpha_i x^i+\sum_{i=1}^n \beta_i/x^i}{\gamma_0+\sum_{i=1}^n \gamma_i x^i+\sum_{i=1}^n \delta_i/x^i},
\ee
reads
\be \label{tsuda-map}
\phi: (x^1,\ldots, x^n)\mapsto (X^1,\ldots,  X^n), \;\;\mbox{where} \;\; X^i=\frac{1}{x^i} \frac{\beta_i-\delta_i H(x^1, \ldots , x^n)}{\alpha_i-\gamma_i H(x^1, \ldots , x^n)},\;\; i=1,\ldots, n.
\ee
It is an involution and moreover is a $2^n-$rational map in the  narrow sense i.e. the map the companion maps and their inverses have the same functional form.
\epr

\bpf
By varying the integral (\ref{tsuda-int}), apart the identity solution, the equations
$$
\forall k \in\{1,\ldots n\} \;\;\; H(x^1,\ldots , X^k, \ldots x^n)-H(x^1, \ldots , x^n)=0
$$
has as solutions the involutions
$\sigma^k: (x^1,\ldots , x^k, \ldots x^n)\mapsto (x^1,\ldots , X^k, \ldots x^n),\;\; k\in \{1,\ldots n\}$ where
\be \label{invt}
 X^k=\frac{1}{x^k} \frac{\beta_k-\delta_k H(x^1, \ldots , x^n)}{\alpha_k-\gamma_k H(x^1, \ldots , x^n)}, \;\; k\in \{1,\ldots n\}.
\ee
For any 2 involutions $\sigma^i,\sigma^j,\; i,j \in\{1,\ldots n\}$ of (\ref{invt}), since $H$ is preserved by both of them, it is easy to show by direct computation that  $(\sigma^i \circ\sigma^j)^2=id,\;\; i,j \in\{1,\ldots n\}.$ Hence for the map $\phi=\sigma^1\circ \sigma^2 \circ \ldots \circ \sigma^n,$ there is $\phi^2=id$ so $\phi$ it is an involution  and reads:
$$
\phi: (x^1,\ldots, x^n)\mapsto (X^1,\ldots,  X^n), \;\;\mbox{where} \;\; X^i=\frac{1}{x^i} \frac{\beta_i-\delta_i H(x^1, \ldots , x^n)}{\alpha_i-\gamma_i H(x^1, \ldots , x^n)},\;\; i=1,\ldots, n.
$$
Moreover $\phi$ is a $2^n-$rational map. To prove the $2^n-$rationality, we have to find the companion maps of (\ref{tsuda-map}).
For the coordinates ${\mathcal X}^{m}=(x^1,x^2,\ldots, X^{m_1}, x^{m_1+1},\ldots, X^{m_2},\ldots X^{m_k},\ldots x^n )$ it  holds
 \be \label{tst}
 H(x^1,\ldots x^n)=H(x^1,x^2,\ldots, X^{m_1}, x^{m_1+1},\ldots, X^{m_2},\ldots X^{m_k},\ldots x^n ),
 \ee
 since $H$ is an integral.
 Solving  $k$ of the equalities  of (\ref{tsuda-map}) for $x^{i}, i\in m$  and using (\ref{tst}) we obtain:
\be \label{xis}
x^{i}=\frac{1}{X^{i}}\frac{\beta_i-\delta_i H(x^1,x^2,\ldots, X^{m_1}, x^{m_1+1},\ldots, X^{m_2},\ldots X^{m_k},\ldots x^n )}{\alpha_i-\gamma_i H(x^1,x^2,\ldots, X^{m_1}, x^{m_1+1},\ldots, X^{m_2},\ldots X^{m_k},\ldots x^n )}, \;\;i\in m.
\ee
Where $m$ the set of  $k$ indices (see the notation introduced previously).
 For the remaining $n-k$ equalities of (\ref{tsuda-map}),  we have
\be \label{xjs}
X^j=\frac{1}{x^j}\frac{\beta_j-\delta_j H(x^1,x^2,\ldots, X^{m_1}, x^{m_1+1},\ldots, X^{m_2},\ldots X^{m_k},\ldots x^n )}{\alpha_j-\gamma_j H(x^1,x^2,\ldots, X^{m_1}, x^{m_1+1},\ldots, X^{m_2},\ldots X^{m_k},\ldots x^n )},\;\; j\in m^c.
\ee
So we have the companion maps $\phi^{c_m}: {\mathcal X}^{m}\mapsto {\mathcal X}^{m^c},$ or
$$
\begin{array}{l}
\phi^{c_m}: \; (x^1,x^2,\ldots, X^{m_1}, x^{m_1+1},\ldots, X^{m_2},\ldots X^{m_k},\ldots x^n )\mapsto (X^1,X^2,\ldots, x^{m_1}, X^{m_1+1},\ldots, x^{m_2},\ldots x^{m_k},\ldots X^n ),
\end{array}
$$
where the variables $x^{i},\; i\in m$ are given explicitly in (\ref{xis}), and the variables $X^j,\; j\in m^c$ are given in (\ref{xjs}).

Clearly there are $2^n$ such maps $\phi^{c_m}$. All these maps  have the same functional form as the original map $\phi,$  hence they are involutions. So mapping $\phi$ is a $2^n-$rational map in the  narrow sense.

\epf

\subsection{n=1. The  birational case}
For $n=1$ and if we rename $x^1$ as $x$ we have
\be \label{bi-integral}
H(x)=\frac{\alpha_0+\alpha_1 x+ \beta_1/x}{\gamma_0+\gamma_1 x+ \delta_1/x}
\ee
and the map $\phi$ reads:
\be \label{birational}
\phi: \;\; x \mapsto X=\frac{1}{x} \frac{\beta_1-\delta_1 H(x)}{\alpha_1-\gamma_1 H(x)}=-\frac{\beta_1 \gamma_0-\alpha_0\delta_1+(\beta_1\gamma_1-\alpha_1\delta_1)x}{\beta_1 \gamma_1-\alpha_1\delta_1+(\alpha_0\gamma_1-\alpha_1 \gamma_0)x}.
\ee
Map $\phi$ is a fraction linear involution and as well a $2^1-$rational map (birational). Note that in this case involutivity implies birationality and vice versa.
   Consider the following matrix
${\ds \tau=\left( \begin{array}{lll}
                 \alpha_0& \alpha_1&\beta_1\\
                 \gamma_0& \gamma_1 & \delta_1
                 \end{array} \right).   }$ Let $\tau_{ij}$ the  determinants of the matrix generated by the $ith$ and $jth$ column of $\tau$ (the minor determinants of $\tau$ referred to as Pl\"{u}cker coordinates). Then
  mapping (\ref{birational}) reads ${\ds x\mapsto X=\frac{\tau_{13}-\tau_{32} x}{\tau_{32}-\tau_{12}x}},$  the M\"{obius} involution.


\subsection{n=2. The  quadrirational case}
For $n=2$ if we rename the variables $(x^1,x^2)$ as $(x,y)$ we have
\be \label{quad-rational}
H(x,y)=\frac{\alpha_0+\alpha_1 x+\alpha_2 y+\beta_1/x+\beta_2/y}{\gamma_0+\gamma_1 x+\gamma_2 y+\delta_1/x+\delta_2/y}
\ee
and the map $\phi$ reads:
\be \label{quadrirational}
\phi:(x,y)\mapsto(X,Y)=\left(\frac{1}{x} \frac{\beta_1-\delta_1 H(x,y)}{\alpha_1-\gamma_1 H(x,y)},\frac{1}{y} \frac{\beta_2-\delta_2 H(x,y)}{\alpha_2-\gamma_2 H(x,y)}\right).
\ee
Mapping $\phi$ is involutive and moreover a $2^2-$rational map (quadrirational).  Consider the following matrix
${\ds \tau=\left( \begin{array}{lllll}
                 \alpha_0& \alpha_1&\alpha_2&\beta_1&\beta_2\\
                 \gamma_0& \gamma_1&\gamma_2 & \delta_1&\delta_2
                 \end{array} \right).   }$  As in the previous subsection,  $\tau_{ij}$ denotes the Pl\"{u}cker coordinates. Then  mapping (\ref{quadrirational}) reads:
\be \label{quad-fin}
\phi: \; (x,y)\mapsto \left(X=\frac{\tau_{34}y^2+\tau_{14}y+\tau_{54}-\tau_{42}x y}{\tau_{42}y-(\tau_{23}y^2+\tau_{21}y+\tau_{25})x},Y=\frac{\tau_{52}x^2+\tau_{51}x+\tau_{54}-\tau_{35}x y}{\tau_{35}x-(\tau_{23}x^2+\tau_{13}x+\tau_{43})y} \right).
\ee

\subsection{n=3. The  octarational case}
 For $n=3$ if we rename the variables $(x^1,x^2,x^3)$ as $(x,y,z)$ we have
\be \label{octo-rational}
H(x,y,z)=\frac{\alpha_0+\alpha_1 x+\alpha_2 y+\alpha_3 z+\beta_1/x+\beta_2/y+\beta_3/z}{\gamma_0+\gamma_1 x+\gamma_2 y+\gamma_3 z+\delta_1/x+\delta_2/y+\delta_3/z}
\ee
and the map $\phi$ reads:
\be \label{octorational}
\phi:(x,y,z)\mapsto(X,Y,Z)=\left(\frac{1}{x} \frac{\beta_1-\delta_1 H(x,y,z)}{\alpha_1-\gamma_1 H(x,y,z)},\frac{1}{y} \frac{\beta_2-\delta_2 H(x,y,z)}{\alpha_2-\gamma_2 H(x,y,z)},\frac{1}{z} \frac{\beta_3-\delta_3 H(x,y,z)}{\alpha_3-\gamma_3 H(x,y,z)}\right).
\ee
Mapping $\phi$ is involutive and moreover a $2^3-$rational  map ( octarational).
If we consider the matrix $\tau={\ds \left(\begin{array}{lllllll}
                                          \alpha_0& \alpha_1 & \alpha_2& \alpha_3& \beta_1& \beta_2& \beta_3\\
                                          \gamma_0&\gamma_1&\gamma_2&\gamma_3&\delta_1 &\delta_2&\delta_3
                                    \end{array}\right)},$  mapping (\ref{octorational}), in terms of the determinants $\tau_{ij}$ reads:
$$
\phi:\; (x,y,z)\mapsto  \left(X,Y,Z\right),
$$
where
\be
\begin{array}{l}
X={\ds \frac{\tau_{35}y^2 z+\tau_{45}y z^2+\tau_{15}y z+\tau_{65}z+\tau_{75}y+\tau_{25}x y z}{\tau_{52}y z+(\tau_{32}y^2z+\tau_{42}yz^2+\tau_{12}y z+\tau_{72}y+\tau_{62}z)x}},\\ [3mm]
Y={\ds \frac{\tau_{64}xz^2+\tau_{62}x^2z+\tau_{61}x z+\tau_{67}x+\tau_{65}z+\tau_{63}x y z}
{\tau_{36}x z+(\tau_{34}xz^2+\tau_{32}x^2z+\tau_{31}x z+\tau_{35}z+\tau_{37}x)y}},\\ [3mm]
Z={\ds \frac{\tau_{72}x^2y+\tau_{73}xy^2+\tau_{71}x y+\tau_{75}y+\tau_{76}x+\tau_{74}x y z}
{\tau_{47}x y+(\tau_{42}x^2y+\tau_{43}xy^2+\tau_{41}x y z+\tau_{46}x+\tau_{45}y)z}}.
\end{array}
\ee

\section{Conclusions} \label{sec4}

In this note we present a rich family of $2^n-$rational maps. These maps by construction are involutions and highly symmetric in the sense that the maps and their companion maps have the same functional form.

$2^n-$rational maps, are quite exceptional objects inside the set of rational maps. Although on their own are interesting as mathematical entities we expect to have various connections with discrete integrable systems. It is known  that the  quadrirational maps are related to Yang-Baxter maps and therefore to 2-dimensional difference integrable systems \cite{ABSf,PSTV}. A subclass of   octarational maps    provides solutions to the functional tetrahedron equation \cite{sergeev-1998}. Our next goal is to isolate solutions  of functional  tetrahedron equations within the family of maps (\ref{octorational}).  We also anticipate the connection of $2^n-$rational maps presented here (\ref{tsuda-map}) with solutions of higher {\it simplex} equations \cite{Maillet1989221,MAILLET1989389,dimakis-2015}.

\vspace{1cm}

{\bf Acknowledgements}
P. K and P. D. would like to thank the Cyprus Research Promotion
Foundation for their support through  project number KY-ROY/0713/27.

\end{document}